\documentclass[fleqn,twoside]{article}
\usepackage[headings]{espcrc2}

\readRCS
$Id: espcrc2.tex,v 1.2 2004/02/24 11:22:11 spepping Exp $
\usepackage{graphicx}
\usepackage{epsfig}
\usepackage{graphics}
\usepackage[figuresright]{rotating}

\newcommand{\AmS}{{\protect\the\textfont2
  A\kern-.1667em\lower.5ex\hbox{M}\kern-.125emS}}

\newcommand{\UtwoS}{\ensuremath{\Upsilon(\mathrm{2S})}}
\newcommand{\UthreeS}{\ensuremath{\Upsilon(\mathrm{3S})}}

\title{B-Physics and Quarkonia studies with early ATLAS data}

\author{Erez Etzion  \thanks{erez@cern.ch \newline This research was supported (in part) by the German-Israeli Foundation for Scientific Research and Development and the Israel Science Foundation}
\address[TAU]{Tel Aviv University,
Raymond and Beverly Sackler school of Physics and Astronomy. \\
Tel Aviv 69978, Israel \\}
on behalf of the ATLAS Collaboration \\November 2009.
}
       
\runtitle{B-Physics and Quarkonia studies with early ATLAS data}
\runauthor{E. Etzion}

\begin{document}

\begin{abstract}
 Quarkonia and B-Physics are among the first areas to be investigated with the first data collected by ATLAS.
The ATLAS detector at CERN's LHC is preparing to take data from proton-proton collisions expected to start by the end of 2009.  Investigation of the decay of B-hadrons represents a complementary approach to direct searches for physics beyond the Standard Model. Early B-Physics data will provide valuable information on the detector performance, as well as allow calibration studies in support of New Physics searches. Meaningful quarkonia studies performed with early data are expected to have the reach to make authoritative statements about the underlying production mechanism and provide cross-sections in this new energy regime. We review various aspects of prompt quarkonium production at the LHC: the accessible ranges in transverse momentum and pseudo-rapidity, spin alignment of vector states, separation of color octet and color singlet production mechanism and feasibility of observing radiative  $\chi_c$  decays. 
\vspace{1pc}
\end{abstract}

\maketitle

\section{Introduction and motivations}
\label{chap:intro}

The main goal of ATLAS \cite{ATLAS}, one of the general purpose Large Hadron Collider (LHC) experiments, is looking for signals of New Physics beyond the Standard Model (BSM). While the expected BSM signatures are mostly at high transverse momenta ($p_T$), the B-Physics processes are concentrated in the low $p_T$ region. Nevertheless ATLAS is able to provide B-Physics measurements based on its precise vertexing and tracking, its muon identification and a dedicated trigger strategy. 
The LHC is expected to produce a large number of B decays and heavy quarkonia states such as J/$\psi$ and  $\Upsilon$  in the low luminosity runs during the first few years of running.  Their sizable branching fraction into charged lepton pairs allows for easy separation of these events from the expected huge hadronic background. 
The ATLAS efforts concentrate on those B decays that can be selected by the first and second trigger levels. The most favorable B and heavy quarkonium trigger signatures will be events decaying to $\mu$ pair, either directly or via a J/$\psi$ meson. 
Using the J/$\psi$ trigger ATLAS will be able to collect unprecedentedly high statistics of $B_s\rightarrow J\psi\phi$ decays, allowing measurements of CP violating effects, which are predicted by some BSM models to be significantly larger than the SM prediction. The di-muon trigger will also give ATLAS access to potentially large numbers of rare $B_s\rightarrow\mu\mu$ decays. Exclusive decay channels of $B^+$ and $B^0_d$ with two muons in the final state, will be measured for studying mass and lifetime as well as production cross-section.
Being narrow resonances, the quarkonia are perfectly suited for alignment and calibration of the ATLAS trigger and tracking systems.
On top of that, understanding the details of the prompt charmonium production is a challenging task and a good testbed for various QCD calculations, spanning both perturbative and non-perturbative regimes through the creation of heavy quarks in the hard process and their subsequent evolution
into physical bound states.

\section{ATLAS B-Physics di-muon trigger}
\label{sec:trigger}
\begin{figure}[htb]
\vspace{-2 pt}
\includegraphics[width=8.2cm]{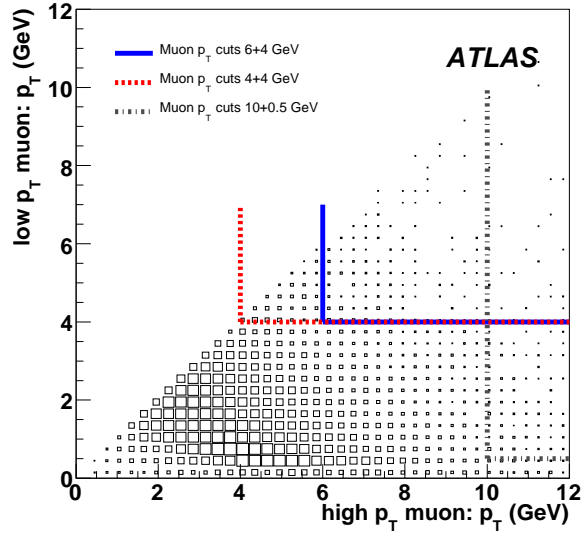}
\caption{Density of J/$\psi$ production cross-section as a function of the hardest and 
      softer muon $p_T$ of muons coming from J/$\psi$. No cut was placed on the generated sample, 
      but the overlaid lines represent the thresholds of observed events with trigger cuts applied.}
\label{fig:Jpsilowpttrigger}
\end{figure}

ATLAS uses a fast and efficient muon based B-Physics triggering strategy. The LHC will collide two 7~TeV proton beams at a rate of 40~MHz, which together with pile-up will result in an interaction rate of up to around 1~GHz. The ATLAS trigger acts to reduce this rate to around 100~Hz in three successive levels, whilst keeping only events `of interest'.
The trigger first level (level-1) decision is based on  coarse granularity of two sub-detector systems: the muon trigger chambers and the calorimeters. 
The second level trigger (level-2) receives from level-1 data restricted to limited Regions of Interest (RoI). For a level-1 muon, the level-2 will use the information from the muon and inner detector tracking chambers to improve the muon momentum estimate, which allows a tighter selection based on this quantity.
There are two specific types of trigger dedicated to quarkonium: one which requires two level-1 RoIs corresponding to two muon candidates 
with $p_T$ values above thresholds of 4 and 6~GeV respectively, and the other requires a single level-1 RoI above a threshold of 4~GeV and searches for the second muon of opposite charge in a wide RoI at level-2. A complementary trigger on J/$\psi$ events is triggering on a single higher $p_T$ (10~GeV) muon and searching for a matched track at the off-line analysis level. Before incorporating trigger and reconstruction efficiencies, the predicted cross-sections
for $pp\rightarrow J/\psi\rightarrow\mu^+\mu^-+X$ were calculated for a number of $p_T$ thresholds on the di-muon trigger (see Table~\ref{tab:oniaxsecs}
for details). Figure~\ref{fig:Jpsilowpttrigger} illustrates the distribution of
cross-sections across the values of the $p_T$ of the harder and softer muon from the quarkonium decay
without any muon cuts applied and zero polarization. 
The lines overlaid on the plots represent the muon trigger thresholds $p_T>6$~GeV and $p_T>4$~GeV for the harder and the software muon respectively (denoted further by "6+4~GeV" or  $\mu6\mu4$).
Similarly "4+4~GeV" ($\mu4\mu4$)  refers to the trigger threshold $p_T>4$~GeV applied on both muons. The "10+0.5~GeV" ($\mu10$) refers to trigger $p_T$ threshold of 10~GeV applied on one of the muons only. In all cases the pseudorapidity, $\eta$, of a muon lies within an interval $|\eta|<2.5$.

Even if the bulk of the J/$\psi$ are produced under the trigger thresholds, nevertheless, due to the high cross-section the number of accepted J/$\psi$ will be substantially larger than at the Tevatron.
In the $\Upsilon$ case, due to it's relatively larger mass the bulk of the production is in the region near the muon threshold of 5 and 4~GeV.
This means that by lowering the threshold for the higher momentum muon we can significantly increase the number of recorded $\Upsilon$ events.

Estimated quarkonia cross-sections for the three scenarios are presented in Table~\ref{tab:oniaxsecs}. The production will be dominated by J/$\psi$ and $\Upsilon(1S)$. 
while $\psi^\prime$ and $\Upsilon(2S)$, $\Upsilon(3S)$ are expected to give smaller contributions.

\begin{table}[htb]
\caption{Predicted cross sections for various quarkonium states into muons, for different pt thresholds (before trigger and reconstruction efficiencies).
  The last column shows the overlap between the di-muon and single muon samples.}
  \label{tab:oniaxsecs}
\begin{tabular}{|c||c|c|c|| c |}
\hline
{Quarkonium}  & \multicolumn{4}{c|}{Cross-section, nb} \\
                             & $\mu4\mu4$ & $\mu6\mu4$ & $\mu10$ & $\mu6\mu4$\\
                              &           &            &          & and $\mu10$ \\
\hline
pp$\rightarrow$J/$\psi$           & 28  & 23  & 23   &  5   \\
pp$\rightarrow \psi^\prime$   & 1.0 & 0.8 & 0.8  &  0.2    \\
pp$\rightarrow \Upsilon(1S)$          & 48  & 5.2 & 2.8  &  0.8 \\
pp$\rightarrow$ \UtwoS          & 16  & 1.7 & 0.9  &  0.3    \\
pp$\rightarrow$ \UthreeS        & 9.0 & 1.0 & 0.6  &  0.2   \\
\hline
\end{tabular}
\end{table}

Precise cross-section measurements require a good understanding of the event and trigger selection efficiency. To evaluate the trigger efficiency we use tag-and-probe (TAP) method applied on the real $J/\psi\rightarrow\mu^+\mu^-$ data. In the TAP method  a single triggered muon from a reconstructed di-muon decay of an identified specific particle is used as a tagged muon while we probe the trigger efficiency of the second muon. We verified that the efficiency derived with the TAP method on simulated events is in good agreement with the single muon trigger efficiency  directly estimated from Monte Carlo samples.

\section{B-Physics early measurements}
\label{sec:bphys}
The large production cross section of $b\bar{b}$ events will enable, even during the initial low luminosity phase of the LHC, the observation of various exclusive channels such as 
 $B^+\rightarrow J/\psi K^+$, $B_d\rightarrow J/\psi K^0$  or  $B_s\rightarrow J/\psi\phi$.
The measurement of the known properties of the $B$ hadrons such as masses and lifetime will be used to study and  tune the detector. The known channels will serve as a reference while searching for the rare $B$ decays.

The charmonium events and the $B$ decays to $J/\psi$ share a similar detection strategy.
In any event which passes the di-muon trigger, all the reconstructed muon candidates are combined into oppositely charged pairs, 
and each of these pairs is analyzed in turn. The invariant mass of the combined muons is required to be within a defined window around the $J/\psi$ mass. In the direct charmonium search, if the invariant mass of the two muons is above 1~GeV, we first attempt to refit the tracks to a common vertex.
If a good vertex fit is achieved, the pair is accepted for further analysis. 
If the invariant mass of the refitted tracks is within 300~MeV for J/$\psi$ or 1~GeV for $\Upsilon$ (six times the expected average mass resolution) of the expected mass, the pair is considered as a charmonium candidate.

\begin{figure}[htbp]
  \includegraphics[width=8cm]{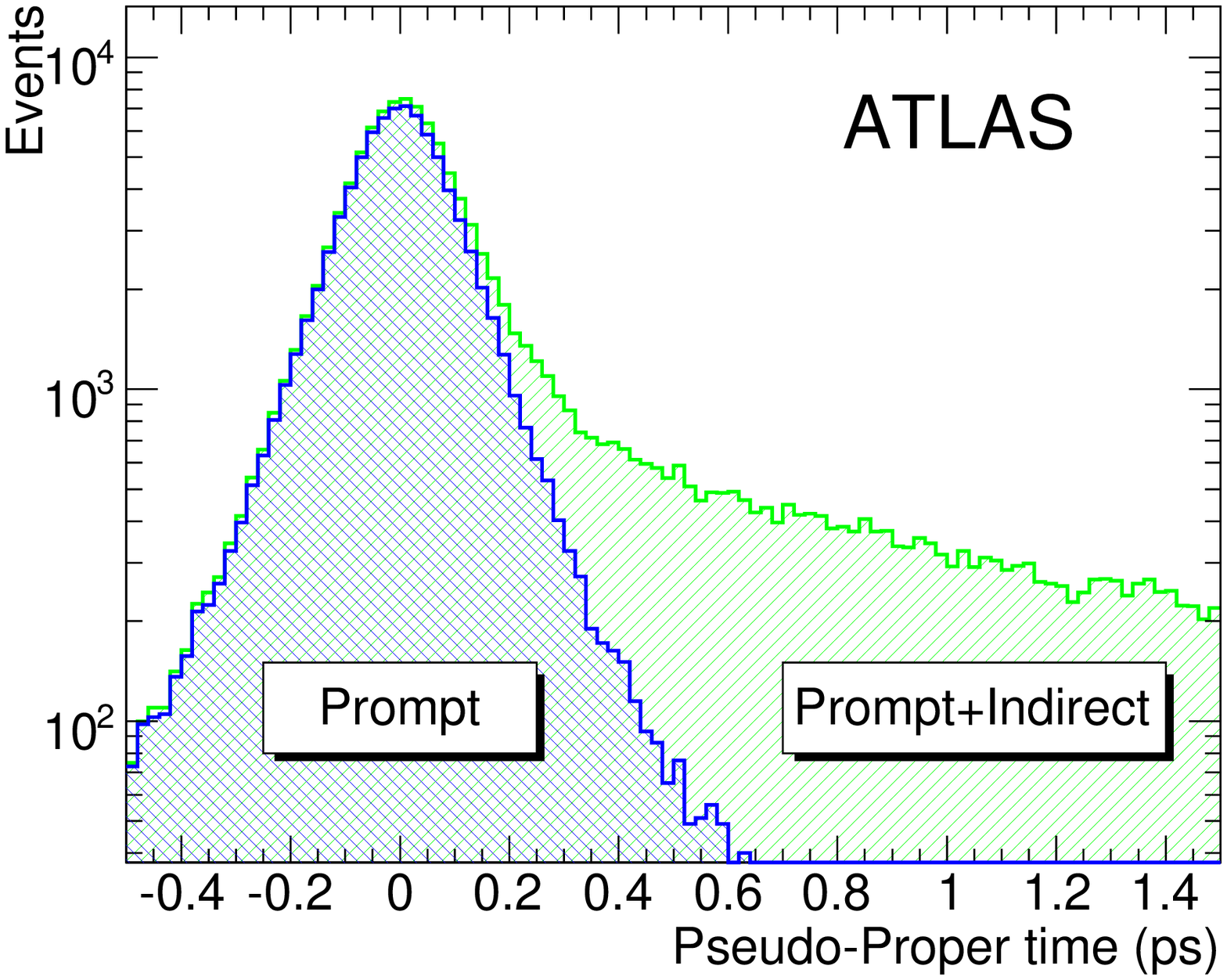}
  \includegraphics[width=8cm]{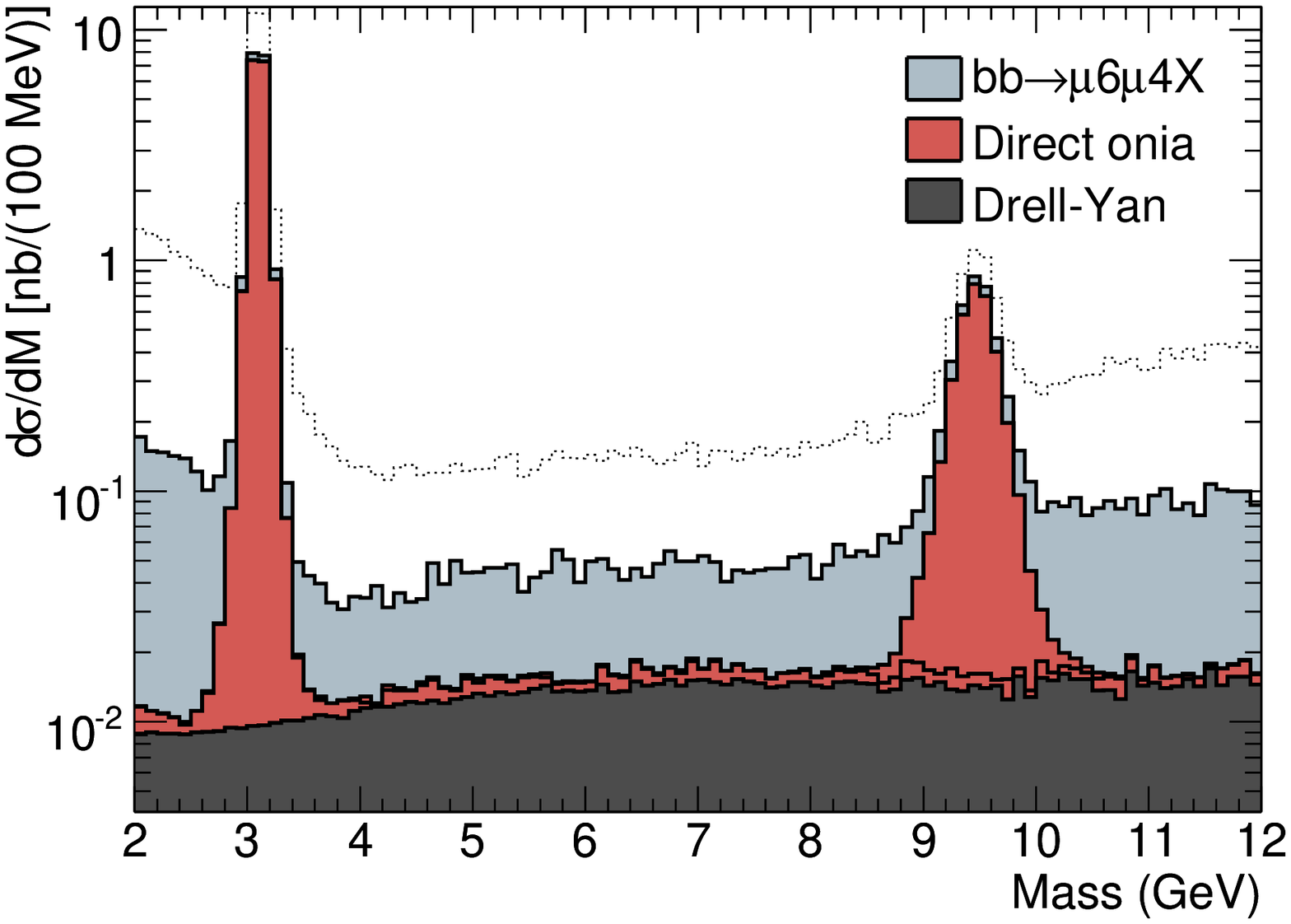}
\hfill\\
  \caption{Upper - Pseudo-proper decay time, $\tau$,  distribution for reconstructed prompt J/$\psi$
      (cross-hatched, centred at zero) and the sum of prompt and indirect J/$\psi$ from B-decays
      (hatched, exponential distribution).\newline
      Lower -  Sources of low invariant mass di-muons, reconstructed
      with a $\mu6\mu4$ $p_T$ trigger, with the requirement
      that both muons are identified as coming from a primary vertex
      and with a pseudo-proper decay time smaller than 0.2~ps. The white area represents the background that was rejected with the pseudo-proper decay time treatment.}
    \label{fig:jpsipropertime}
\end{figure}

The radial displacement of the two-track vertex from the beamline is used to distinguish
between prompt J/$\psi$ originated at the proton proton interaction and the B-hadron decays into quarkonia.  The pseudo-proper decay time, $\tau$, is defined as
\begin{center}
\begin{equation}
\tau = \frac{L_{xy}\cdot M_{J/\psi}}{p_T(J/\psi)},
\end{equation}
\end{center}
where $M_{J/\psi}$ and $p_T(J/\psi)$ represent the J/$\psi$  invariant mass and
transverse momentum, and $L_{xy}$
is the transverse decay length of the meson. The resolution in the pseudo-proper decay time is expected to vary from 0.110~ps for the low $p_T$ charmonia down to 0.07~ps for the higher $p_T$. 
As demonstrated in  Figure~\ref{fig:jpsipropertime} a cut on this quantity can efficiently distinguish between prompt and indirect J/$\psi$ events.

In the case of $B^+\rightarrow J/\psi K^+$ all tracks with positive charge which do not originate from the primary vertex are considered as $K^+$.  We fit the three tracks corresponding to a $J/\psi$ and the $K^+$ candidates  to a common vertex. The momentum sum of the two has to point to the primary vertex. A maximum likelihood (ML) method is used to fit the $B^+$ mass distribution taking Gaussian probability density function for the signal and linear distribution for the background. The mass resolution in this channel is estimated, for an integrated luminosity of 10 $pb^{-1}$ to be $\sigma=42.2\pm 1.3$ MeV with a total efficiency of $29.8\pm 0.8\%$. The differential cross-section is obtained in this channel in different bins of  $p_T(B^+)$ using the number of the signal events and the efficiency from the ML fit to the invariant mass distribution. The total cross-section is determined with the same procedure but for the whole $p_T$ region of $B^+$. For an integrated luminosity of  10 $pb^{-1}$, we determined that the total production cross section can be measured with a statistical precision better than 5\% and the differential cross-section can reach a precision of 10\%. The systematic uncertainties are expected to be dominated by the luminosity and the branching ratio.

The channel  $B_s\rightarrow J/\psi\phi$ has the potential to reveal New Physics signals. The weak mixing phase $\phi_s$, which in the SM is very small, may be enhanced by BSM processes. The time dependent angular distribution of this decay depends on seven physics parameters: two independent complex transversely amplitudes, the mean lifetime and the mass eigenstate width difference ($\Delta\Gamma$), and the weak mixing phase.
The full analysis will require a fit to these parameters and is sensitive to statistics, experimental resolution of lifetime, mass and decay angles as well as flavour tagging performance and background rejection.
The experience of the Tevatron experiments taught us that a few thousands events only allow a two dimensional profile likelihood fit in $\phi_s$-$\Delta\Gamma$ plane. The LHC after the high luminosity runs will enable a simultaneous determination of all seven parameters. With early data the ATLAS program will begin with calibration measurements supporting this analysis. The topological identical channel $B_d\rightarrow J/\psi K^0$,  which is expected to yield 15 times more statistics, is the the primary background but is also an essential as a control sample. It will allow high precision tests of the lifetime measurement as well as flavour tagging analysis calibration. With the earliest data, ATLAS will use a fit to simultaneously access the mean mass and lifetime of the $B_s$ and the $B_d$ mesons. This will enable the testing of our understanding of the tracking at low $p_T$ scale after 150 $pb^{-1}$, and will start to improve the world precisions on these measurements after about 1 $fb^{-1}$. Early on, topologically similar backgrounds will be admitted to the analysis and secondary vertex cuts will not be applied. After 10 $fb^{-1}$ the precision on the $B_d$ lifetime is expected to be 10\% and similar precision for the $B_s$ mean lifetime is expected after 150 $pb^{-1}$.

\section{Charmonium analysis}
\label{sec:charmonium}

Charmonium production comprises of three main processes: direct singlet production, octet production and singlet production of $\chi$
states. Each of these processes is characterized by different differential cross-sections, $\frac{d\sigma}{dp_T}$.

The reconstruction efficiency of charmonium varies with $p_T$ and $\eta$. When the $p_T( J/\psi)>10$~GeV we get a sharp rise in its acceptance. The rise in the $\Upsilon$ case is less sharp and it reaches a high acceptance level around 30~GeV. Both channels reach a similar plateau at acceptance of around 85\%. The trigger requires both muons to pass the $p_T$ threshold and so the angular separation between the two muons is not large for most of the accepted events. We describe the opening angle by $\delta R=\sqrt{\delta\phi^2+\delta\eta^2}$, where $\delta\phi$ and $\delta\eta$ are the differences in the azimuthal angle and the  pseudorapidity of the two muons from J/$\psi$.  Typical values for $\delta R$ are $\approx 0.47$ which means that requiring both muons to be above the $p_T$ threshold forces them to fly very close one to each other.
For this reason the J/$\psi$ angular acceptance follows closely the individual muon distribution and its dependence on material and detector effects. In the $\Upsilon$ case, due to its higher mass, it tends to be produced with higher $p_T$. Therefore the separation angle $\delta R$ is much wider, the muons do not go to the same area in the detector and  the efficiency dependence on pseudorapidity is much smoother than in the J/$\psi$ case.
The main sources of low invariant mass di-muons expected to dominate the background for prompt charmonium are: decays in flight of $\pi^\pm$ and $K^\pm$ (which are reduced after cuts to below the 1\% level); di-muon production via the Drell-Yan process (which are rendered essentially negligible by the trigger requirements); continuum of muon pairs from beauty (and charm) decays and the main source - indirect J/$\psi$ production - the reduction of which is discussed below.

All the background sources apart from the Drell-Yan pairs contain muons which do not originate from the interaction point, and this is used to suppress their contamination by rejection of the events containing a secondary vertex if identified.
As detailed in the previous section a cut on the pseudo-proper decay time is used to distinguish between the direct and indirect charmonium cases. Figure ~\ref{fig:jpsipropertime}  shows the pseudo-proper decay time distribution of the two samples and the impact on the mass distribution when rejecting events with a pseudo-proper time above 0.2~ps.

Using the $\mu10$ trigger each reconstructed single muon candidate
is combined with oppositely-charged tracks reconstructed in the same event within a cone of $ \delta R = 3.0$. Any track, including those that were not identified as muons, are examined.
As in the di-muon analysis, we require that both the identified muon and the
track are flagged as having come from the primary vertex. In addition,
 we impose a cut on the transverse impact parameter $d_0$, 
$|d_0|<0.04$~mm on the identified muon and $|d_0|<0.10$~mm on the second track, in order to further suppress 
the number of background pairs from $B$-decays.
We obtain a J/$\psi$  invariant mass resolution close to that in those events passing the trigger requiring both muons to be confirmed.
It's worth noting that the signal-to-background ratio
around the J/$\psi$ peak improves slightly with increasing transverse
momentum of J/$\psi$.
At higher $p_T$ the $\cos\theta^\ast$ acceptance also becomes 
broader, which should help independent polarization measurements.

\begin{figure}[htb]
\vspace{-2 pt}
\includegraphics[width=8.2cm]{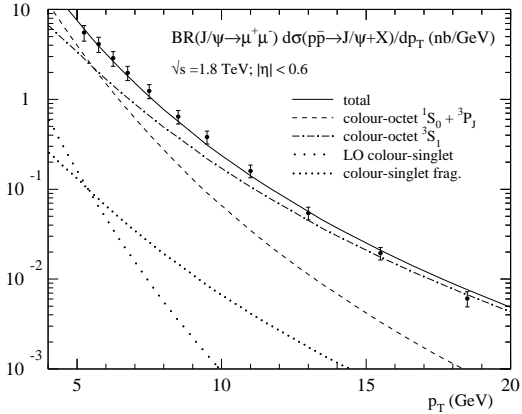}
\caption{Differential cross-section of J/$\psi$ production at CDF with theoretical predictions for color-singlet and color-octet model production\,\cite{kramer-cdf}}
\label{CDF-cross}
\end{figure}

\begin{figure}[htb]
\vspace{-2 pt}
\includegraphics[width=7.8cm]{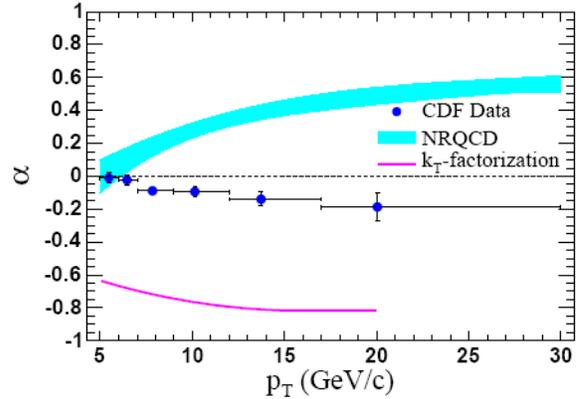}
\caption{Polarization of J/$\psi$  production as a function of $p_T$ at CDF, (data points), 
curves for limit cases of the $k_T$ factorization model,
      and a band for NRQCD predictions (from \,\cite{Abulencia:2007us}).}
\label{CDF-pol}
\end{figure}

ATLAS is capable of detailed checks of the predictions of various models describing the evolution of heavy quark antiquark pair into a quarkonium bound state. Among the models which are considered are: Color Evaporation Model (CEM)\,\cite{Amundson-all}, the Color Singlet Model (CSM)\,\cite{orig-ccbar} which failed to predict the J/$\psi$ production rate measured by CDF,
and the Nonrelativistic QCD (NRQCD) Colour Octet Model (COM)\,\cite{Bodwin-Braaten} which was proposed to explain this discrepancy. The good description of the Tevatron data by the COM model shown in Figure~\ref{CDF-cross} is at least in part due to the tuning of some of its parameters which were determined from the same data. However, the difference between the CDF measurement of J/$\psi$ polarization dependence on transverse momentum ($p_T$) and the theory predictions (Figure~\ref{CDF-pol}) is motivating us to repeat the measurement in the LHC's higher energy and luminosity regime.

The polarization of  quarkonia can be measured using the angular distribution 
of the daughter particles produced in the decay. 
\begin{center}
\begin{equation}
\label{eqn:spinalign}
  \frac{d\Gamma}{d\cos\theta^\star}\propto 1+\alpha \cos^2 \theta^\star,
\end{equation}
\end{center}
where  $\cos\theta^\star$ is the angle between the direction
of the positive (by convention) muon from 
quarkonium decay in the quarkonium rest frame, and the direction of the quarkonium itself in the laboratory
frame.
The polarization parameter $\alpha$, defined as 
$\alpha=(\sigma_T-2\sigma_L)/(\sigma_T+2\sigma_L)$, is equal to $+1$ for 
transversely polarized quarkonia production (helicity $\pm1$). $\sigma_T$ and $\sigma_L$ are the transverse and longitudinal cross sections. For longitudinal (helicity 0) polarization, $\alpha$ is equal 
to $-1$. Unpolarized production consists of equal fractions of helicity states 
$+1$, $0$ and $-1$, and corresponds to $\alpha=0$. 

The previous Tevatron measurements were limited to below around 20~GeV,
where the polarization is best predicted and the theory most understood.
At ATLAS we aim to measure the polarization of directly produced 
prompt quarkonium in the $p_T$ region up to
$\sim 50$~GeV and beyond with extended coverage in $\cos\theta^\star$. This
will allow for improved fidelity of efficiency,
provide better discrimination of longitudinal and transverse polarizations and therefore reduced systematic uncertainties.


%



\begin{figure*}[h]
\begin{center}
     \includegraphics[width=12.0cm]{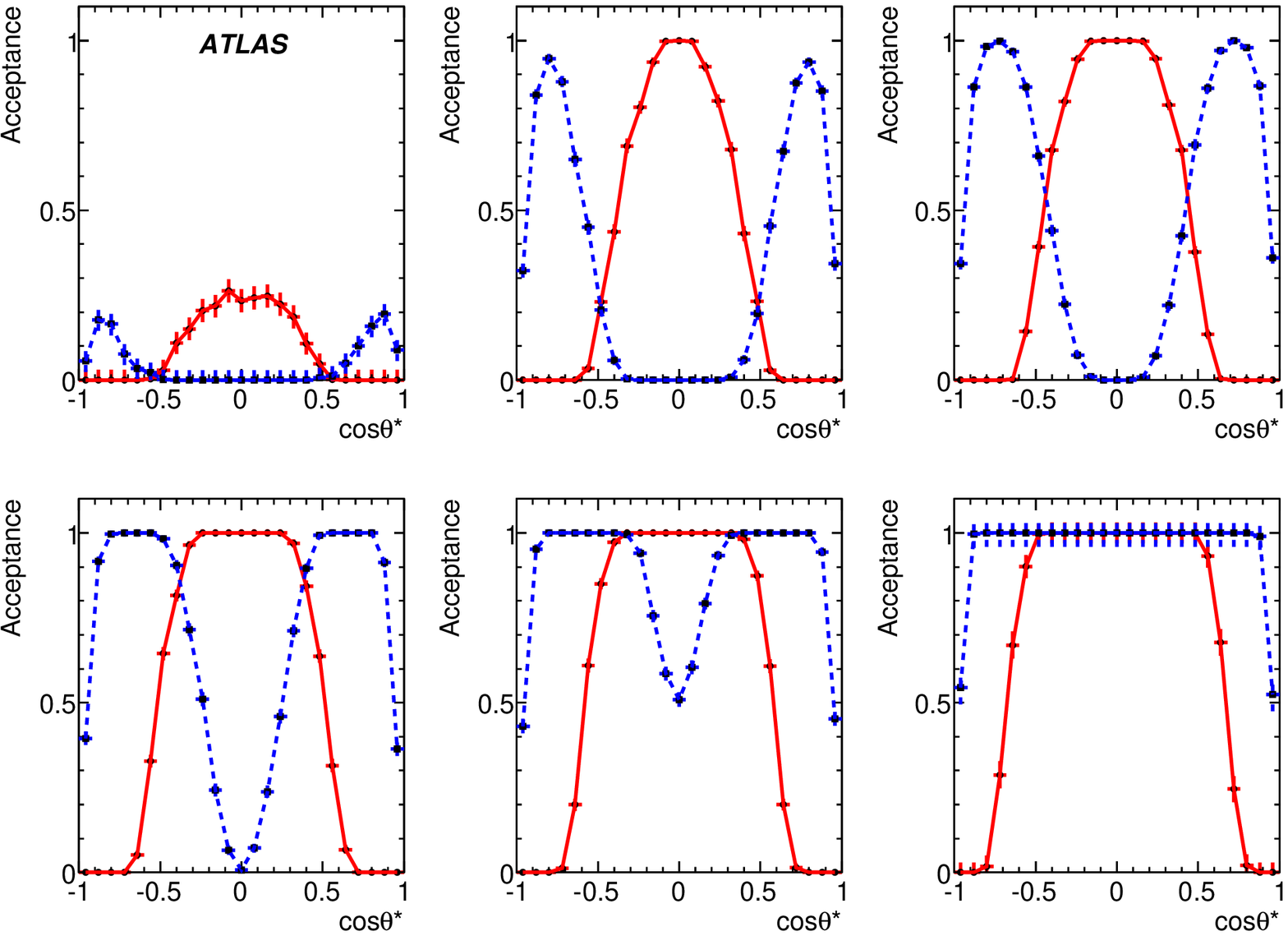}
     \includegraphics[width=12.0cm]{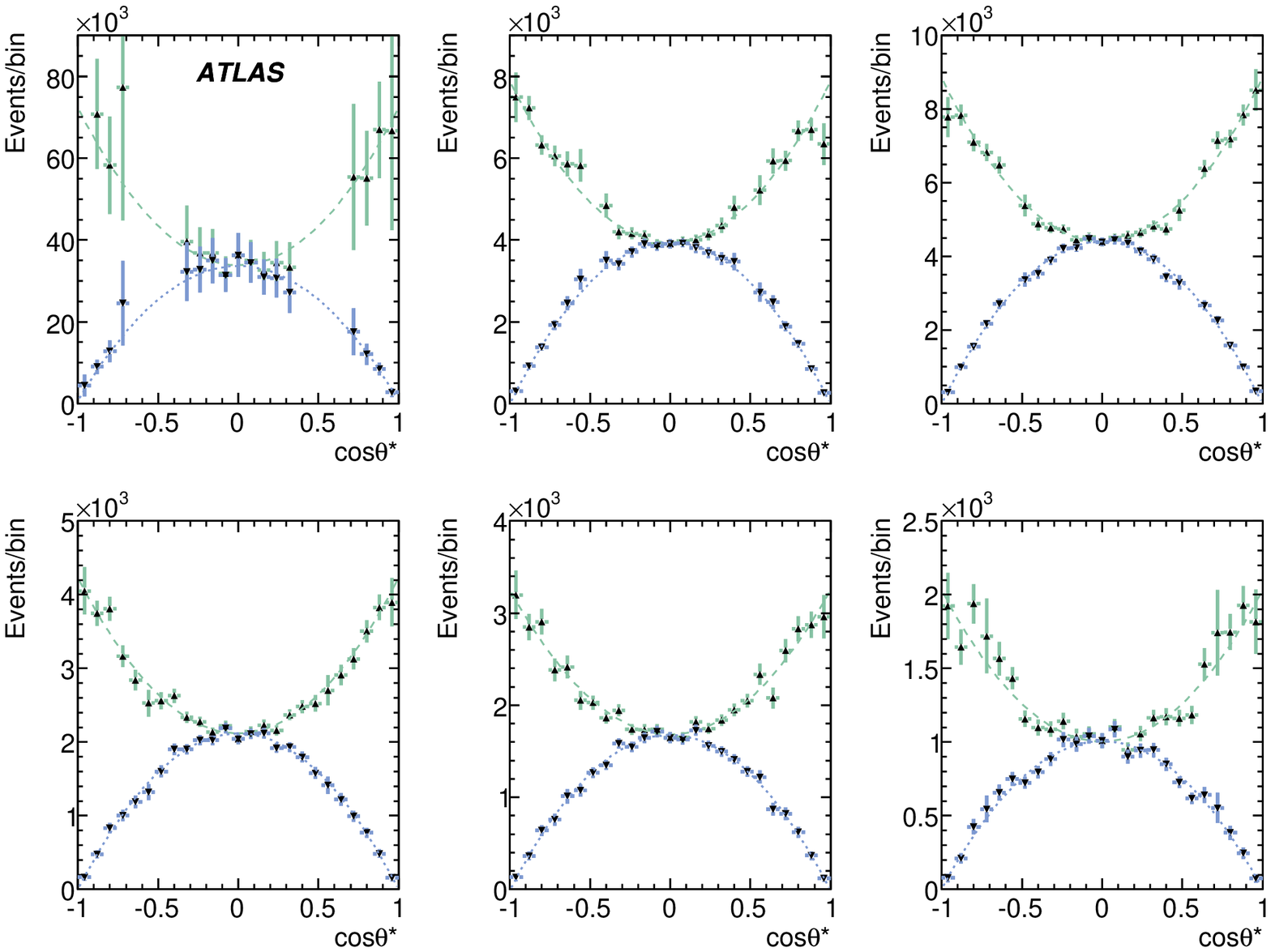}
\hfill\\ 
\end{center}
   \caption{Six figures at the top show kinematic acceptances of the $\mu6\mu4$ (solid line)
     and $\mu10$ (dashed lines) generator level cuts, 
     calculated with respect to the sample with no generator level 
     cuts on muon $p_T$, in slices
     of J/$\psi$ transverse momentum: left to right, top to bottom 9-12
     12-13, 13-15, 15-17, 17-21, above 21~GeV.\newline
     Six bottom figures show combined and corrected (equation ~\ref{eq:measure}) distributions in polarization angle, for longitudinally (dotted lines) and transversely (dashed lines) polarized J/$\psi$ sample in the same $p_T$ slices with statistics corresponding to an integrated luminosity of  10~pb  $^{-1}$.}
      \label{fig:jpsimu6mu4-polarisation-ptslices}
\end{figure*}

Evidently the polarization measurement can significantly suffer from low $|\cos\theta^\star|$ acceptance, and hence from difficulties 
in separating detector efficiency corrections from polarization state effects.
The di-muon trigger requirement for both muons to be above a certain $p_T$ threshold ($\mu6\mu4$) reduces to a minimum the acceptance at 
large values of $|\cos\theta^\star|$, where the  difference between various polarization states is the more pronounced (see for example the two polarization states on the lower plots of Figure~\ref{fig:jpsimu6mu4-polarisation-ptslices}).
The acceptance of the single muon trigger sample ($\mu10$) is very different. Here the efficiency is higher at large values of $|\cos\theta^\star|$ and drops  around zero. Figure~\ref{fig:jpsimu6mu4-polarisation-ptslices} (on the top) demonstrates how the two samples complement each other mainly at the low $p_T$ regions, while at high $p_T$ the two triggers increasingly overlap, thus allowing for a cross-check of acceptance and efficiency corrections.
The $p_T$ distributions of both samples, $\mu6\mu4$ and $\mu10$ events,
were appropriately  combined.
The combined distribution $dN^{\mathrm{raw}}/d\cos\theta^\ast$, 
was corrected according to equation~\ref{eq:measure}, 
\begin{eqnarray}
\frac{dN^{\mathrm{cor}}}{d\cos\theta^\ast} =
\frac{1}{{\cal{A}}(p_T,\cos\theta^\ast)
\cdot \varepsilon_1
\cdot \varepsilon_2}
\cdot\;
\frac{dN^{\mathrm{raw}}}{d\cos\theta^\ast}.
\label{eq:measure}
\end{eqnarray}
Here $\varepsilon_1$ stands for the trigger and reconstruction
efficiency, while $\varepsilon_2$ denotes the efficiency of
background suppression cuts for each sample, and ${\cal{A}}(p_T,\cos\theta^\ast)$ is the kinematic acceptance of the trigger selection.
The results obtained when fitting an unpolarized sample ($\alpha=0$) with statistics corresponding to 10~pb$^{-1}$ are:
$\alpha=0.156\pm0.166, -0.006\pm0.032, 0.004\pm0.029, -0.003\pm0.037, -0.039\pm0.038$ and $0.019\pm0.057$ corresponding to the $p_T$ bins as in Figure~\ref{fig:jpsimu6mu4-polarisation-ptslices}. The precision of the cross-section derived from the normalization factor are $\pm 4.35$ in the first slice and decreasing from $\pm0.09$ to $\pm0.04$ with increasing $p_T$ slices. Repeating the same study with $\Upsilon$ we get back numbers that are consistent with ($\alpha=0$) but with larger errors running from $\pm0.17$ to $\pm0.22$.
To further check the ability to measure the polarization we reweighted our raw Monte Carlo (MC) distribution to emulate a transversely polarized ($\alpha=+1$) and a longitudinally polarized ($\alpha=-1$) J/$\psi$ samples. The same analysis was repeated and the results can be seen on the bottom of Figure~\ref{fig:jpsimu6mu4-polarisation-ptslices}.

A complementary method for polarization measurement has been examined as well
\cite{etzion}. 
 Here a linear combination of three MC generated samples representing longitudinally polarized, transversely polarized and unpolarized events were fitted to the angular distribution of the data. %
The measured values of $\alpha$ in bins of $p_T$ 
were well concentrated around the initial value of the polarization 
that was set in the MC-data sample.

\section{Analysis of $\chi$ production}
\label{sec:chireco}

A sizeable fraction of prompt J/$\psi$ and $\Upsilon$ 
are expected to originate from radiative decays of heavier states,
$\chi_c$ and $\chi_b$.
These states have even $C$ parity and therefore have a strong coupling to
the color-singlet two gluon state.
About 30 to 40\% of J/$\psi$ in our signal will come from decays of $\chi_c \rightarrow J/\psi +\gamma$.
Unfortunately, the energies of these photons tend to be quite small,
and the ability of ATLAS to detect these photons and resolve various $\chi$
states is rather limited.
For  $\chi_c$ reconstruction, 
each selected quarkonium candidate is combined with every reconstructed and identified photon candidate in the event, and the invariant mass of the $\mu\mu\gamma$ system is calculated.
The  $\mu\mu\gamma$ system is considered to be a $\chi$ candidate, if
the difference between the invariant masses of the $\mu\mu\gamma$ and
$\mu\mu$ systems lies between 200 and 800~MeV, and
the cosine of the opening angle between the
J/$\psi$ and $\gamma$ is larger than 0.97.

We fitted three guassians to the difference in invariant masses of the $\mu\mu\gamma$ and
$\mu\mu$ measured in those $\chi$ candidates. 
%
The MC input amplitudes of the peaks 
were reproduced reasonably well \cite{etzion}.
The overall $\chi_c$ reconstruction efficiency is estimated to be about 4\%.

\section{Physics reach with early ATLAS data}
\label{sec:summary}
The ATLAS B-Physics program will run from the earliest days 
and will pursue indirect searches for New Physics via B-hadrons decays.
An efficient, fast and clean di-muon trigger scheme will allow ATLAS to collect large samples of B-hadrons and quarkonia throughout the lifetime of the experiment. 
Reconstruction of exclusive B-meson decays with two muons in the final state, especially with $J/\psi$, is foreseen already with the first $pb^{-1}$. 
Early B-Physics data will provide valuable information on the detector performance and will  allow calibration studies in support of New Physics searches.
Meaningful quarkonium studies performed with early data are expected to have a sufficient reach to make authoritative statements about the underlying production mechanism and provide cross-section measurements in this new energy regime.
Quarkonium spin-alignment measurements at ATLAS will have the capability to distinguish between various production models of quarkonium.  
Even in the early stages ATLAS is expected to provide competitive measurements  of the  $J/\psi$ and 
$\Upsilon$ polarization - $p_T$ dependent values. 
With several million J/$\psi\rightarrow\mu\mu$ decays, and better understanding
of the detector,  $\chi_c\rightarrow J\/\psi\gamma$ should become observable.




\end{document}